\documentclass[12pt]{article}
\usepackage{amsmath,amssymb,amsfonts,mathrsfs, verbatim}
\usepackage{amsmath,amssymb,amsfonts}
\usepackage{bm}% bold math
\usepackage{dcolumn}
\def\d{\partial}
\def\half{\frac{1}{2}}
\def\D{\mathscr{D}}

\def\beq{\begin{equation}}
\def\eeq{\end{equation}}
\def\bea{\begin{eqnarray}}
\def\eea{\end{eqnarray}}

\title{\textbf{Dreibein as prepotential for three-dimensional Yang-Mills theory}}
\author{Indrajit Mitra$^{a}$\footnote{E-mail: indra.theory@gmail.com, imphys@caluniv.ac.in}~ and
H. S. Sharatchandra$^{b}$\footnote{E-mail: sharat@cpres.org} \\\\
$^a$ Department of Physics, University of Calcutta,
92 A.P.C. Road, \\Kolkata 700009, India\\
$^b$ Centre for Promotion of Research,\\
7, Shaktinagar Main Road, Porur, Chennai 600116, India}
\date{}
\begin{document}
\maketitle
\begin{abstract}
We advocate and develop the use of the dreibein (and the metric) as
prepotential for three-dimensional SO(3) Yang-Mills theory.
Since the dreibein transforms homogeneously under gauge transformation,
the metric is gauge invariant. For a generic gauge potential, there is a unique
dreibein on fixing the boundary condition. 
Topologically non-trivial monopole configurations are given by conformally flat metrics, with scalar
fields capturing the monopole centres.
Our approach also provides an ansatz for the gauge potential covering
the topological aspects. 

\end{abstract}


%\tableofcontents
%\newpage
%%%%%%%%%%%%%%%%%%%%%%%%%%%%%%%%%%%%%%%%%%%%%%%%%%%%%%%%%%%%%%%%%%%%%%%%%%%%%%%%%%%%%%%%%%%%%%%%%%%%%%%%%
\section{\bf Introduction}
%%%%%%%%%%%%%%%%%%%%%%%%%%%%%%%%%%%%%%%%%%%%%%%%%%%%%%%%%%%%%%%%%%%%%%%%%%%%%%%%%%%%%%%%%%%%%%%%%%%%%%%%%
In this paper we advocate and develop the use of the dreibein or triad (the 3d version of the vielbein) as
the basic variable or
prepotential for Yang-Mills theory. The first clear proposal to use the dreibein came from 
Haagensen and Johnson \cite{hj} in the context of Hamiltonian formalism in 3+1 dimensions.
(Refs.\ \cite{freedman}-\cite{nojiri} are some
other works on Yang-Mills theory involving various constructions of the metric.)
The defining equation for the dreibein is the condition
for the dreibein to be torsion-free with respect to a connection one-form.
The authors of Ref.\ \cite{hj} claimed that there are zero modes
associated with this equation due to non-uniqueness of the dreibein $e^i_a$ 
corresponding to a given gauge potential $A_i^a$, 
but also gave arguments that the zero modes would not affect their discussion.
Then Haagensen, Johnson and Lam \cite{hjl} followed it up with a deformation of the defining equation for the dreibein,
removing  the deformation at the end. 
Refs.\ \cite{pushan1}, \cite{pushan2}, \cite{diakonov} and \cite{nojiri} have  used the
original defining equation for the dreibein as in Ref.\ \cite{hj} and not the deformation of it.
 In this work also, we use the original defining equation of Ref.\ \cite{hj}. 
We analyse the case of a {\it generic} Yang-Mills potential, satisfying $ {\rm det} B_i^a \neq 0$ 
($B_i^a$ being the non-abelian magnetic field) and argue that
a unique dreibein is obtained by fixing the boundary condition. Consequently, 
 the torsion-free dreibein is a useful prepotential. The calculation of the
Jacobian will be affected by the zero modes present for vanishing non-abelian magnetic field, 
but the Jacobian will not be needed in the present work. 

Our formulation is useful in several ways:
\begin{enumerate}
\item
The metric arising out of the dreibein is gauge invariant, and so the topological properties of
the field configurations which we link to it are also gauge invariant. 
The Yang-Mills action
is like $R^2$ theory, but without the diffeomorphism invariance.
It should be noted that
{\it our interest in the present work is about Yang-Mills theory, not gravity}. 
\item
Our formulation can be useful for a non-perturbative understanding of Yang-Mills theory.
We find that the topological, monopole configurations (which may drive confinement) correspond to
conformally flat metrics. In our earlier works \cite{pushan1} and \cite{hms1}, we showed how to
locate such a configuration to an internal point (the `centre') in a gauge  invariant way.
In this work, the `centres' are
identified as certain points at which the conformal mode (a scalar field) is extremum. 
%{\it scalar} field theory which might by itself exhibit confinement (in contrast to lattice {\it gauge} theories).

\item
 There has been extensive interest in obtaining
an ansatz for the Yang-Mills potential that exhibits the topological aspects (\cite{cho}-\cite{niemi5}).
We propose an ansatz which
has a scalar and a spin-two part in addition to a pure gauge. The scalar part comes from
the conformally flat metric which captures the monopole configurations.

\end{enumerate}

Our techniques can be extended in a straightforward way to SO(4) Yang-Mills theory in 4-Euclidean dimensions.
Haagensen and Johnson \cite{hj} have also addressed generalization to other gauge groups.
Our approach can also be applied to the 3+1-
dimensional Yang-Mills theory in the Hamiltonian formulation,
with the physical states described as  wavefunctionals of the metric.

Ever since the proposal of non-abelian gauge theory, the similarity with 
Einstein gravity has led to extensive work exploring the relationship between
the two.  Examples of this on the gravity side include the
Chern-Simons formulation
of 2+1 gravity \cite{CS}, Ashtekar and loop gravity formulation \cite{Ash} and 
the pure connection formulation
of general relativity \cite{krasnov}.
We have already cited several works which attack gauge theory using gravity (\cite{hj}-\cite{hjl});
a recent work is Ref.\ \cite{guo}.
In spite of all this, 
our approach gives a new way to attack the exotic
features of Yang-Mills theory. 

The paper is arranged as follows. In Sec.\ \ref{sec:2}, the dreibein $e^a_i$ is defined through the 
torsion-free condition,
and the gauge invariance of the metric is emphasized.
In Sec.\ \ref{sec:3}, we analyse the existence and uniqueness of $e_i^a$ for a given gauge potential $A_i^a$.
Sec.\ \ref{sec:4} deals with topological (monopole) field configurations
and Sec.\ \ref{sec:5} presents the gauge potential ansatz. In Sec.\ \ref{sec:6}, we discuss our results.

%%%%%%%%%%%%%%%%%%%%%%%%%%%%%%%%%%%%%%%%%%%%%%%%%%%%%%%%%%%%%%%%%%%%%%%%%%%%%%%%%%%%%%%%%%%%%%%%%%%%%    
\section{\label{sec:2}Dreibein as prepotential}
%%%%%%%%%%%%%%%%%%%%%%%%%%%%%%%%%%%%%%%%%%%%%%%%%%%%%%%%%%%%%%%%%%%%%%%%%%%%%%%%%%%%%%%%%%%%%%%%%%%%%
Consider the set of nine first order partial differential equations (see, for example, Ref.\ \cite{hj})
\begin{eqnarray}
\label{db}
\epsilon_{ijk}(\d_j e^a_k + A^{ab}_j e^b_k) = 0,
\end{eqnarray}
which constitute the `torsion-free condition' for the dreibein $ e^a_i(x) $ ($ i = 1,2,3 $ are the space indices
and $ a = 1, 2, 3 $ are the group indices) with respect to a connection one-form $ A^{ab}_i(x) $. 
The dreibein $ e^a_i $ is viewed as the square root of a metric
\begin{eqnarray}
g_{ij}(x) = e^a_i(x) e^a_j(x).             \label{metric}
\end{eqnarray}
If 
${\rm det}e^a_i(x) \neq 0 $, 
we can 
expand the LHS of (\ref{db}) in the $ e^a_i $ basis and
write 
\begin{eqnarray}
\label{Gamma}
\d_j e^a_k + A^{ab}_j e^b_k = \Gamma^l_{jk}e^a_l.
\end{eqnarray}
 Imposing the symmetry
\begin{eqnarray}
\Gamma^l_{jk}= \Gamma^l_{kj}
\end{eqnarray}
is equivalent to the torsion-free condition (\ref{db}). Define
\begin{eqnarray}
\D_i e^a_j = \d_i e^a_j + A^{ab}_i e^b_j - \Gamma^l_{ij}e^a_l.
\end{eqnarray}
Then equation (\ref{Gamma}) is $\D_j e^a_k =0$, and so
$[\D_i, \D_j]e^a_k = 0$. This gives
\begin{eqnarray}
{R^l}_{kij}e^a_l - F^{ab}_{ij}e^b_k = 0              \label{riemann}
\end{eqnarray}
where ${R^l}_{kij}$ is the Riemann tensor and
\begin{eqnarray}
F^{ab}_{ij} = \d_i A^{ab}_j - \d_j A^{ab}_i + \left[A_i, A_j\right]^{ab}\,.    \label{5a}
\end{eqnarray}
(See, for example, Ref.\ \cite{hooft}.) 
Using the inverse matrix $ \{e^{bk}\} $:
\begin{eqnarray}
e^a_k e^{bk} = \delta^{ab},
\end{eqnarray}
we can write
\begin{eqnarray}
 F^{ab}_{ij} =  {R^l}_{kij}e^a_l e^{bk}.                          \label{7}
\end{eqnarray}

It may be helpful to link the above with Cartan's structure equations of general relativity.
Cartan's first equation 
\bea
de^a+{\omega^a}_b\wedge e^b=0\,,
\eea
is our eq.\ (\ref{db});
we have used the symbol $A$ in the place of $\omega$ and and put all  group indices as
upper indices. (Since we have SO(3) and not the Lorentz group, there is no need to distinguish
between upper and lower indices.) Cartan's other equation 
\bea
{R^a}_b=d{\omega^a}_b + {\omega^a}_c \wedge {\omega^c}_b
\eea
in components is (on using the dreibein to convert the group indices of ${R^a}_{bij}$ to space indices)
\bea
\label{Cartan2}
{R^l}_{kij} {e^a_l}{e^{bk}}=\partial_i {{\omega_j}^a}_b-\partial_j {{\omega_i}^a}_b+{[\omega_i,\omega_j]^a}_b\,.
\eea
Equation (\ref{Cartan2}) is equations (\ref{5a}) and (\ref{7}) combined (in our notation).

Eq.\ (\ref{db}) can be written as
\bea
\epsilon_{ijk}(D_j e_k)^a=0,                 \label{db2}
\eea
where $D_j$ is the gauge-covariant derivative:
\bea
D_j^{ab}=\delta^{ab}\partial_j+A_j^{ab}\,.
\eea
Now write
\bea
A^{ab}_i =-\epsilon^{abc} A^c_i\,,
\eea
where $A^c_i$ is the Yang-Mills potential. Then eq.\ (\ref{5a}) gives the Yang-Mills field strength.
$A_i^a$ transforms {\it inhomogeneously} under an
SO(3) gauge transformation, but eq.\ (\ref{db2}) ensures that the dreibein $e_i^a$
transforms {\it homogeneously}:
\begin{eqnarray}
{e^a_i}'(x)= O^{ab}(x)e^b_i(x)\,.            \label{hom}
\end{eqnarray}
Here $O^{ab}$ is an $SO(3)$ matrix. The metric $g_{ij}(x)$ as given by (\ref{metric}) is therefore
{\it gauge-invariant}.
All gauge invariant objects can be rewritten in terms of $ g_{ij} $. For instance,
 \begin{eqnarray}
{\rm tr} \left(F_{i_1j_1}F_{i_2j_2}\cdots\right) = {\rm tr}\left(R_{(i_1j_1)}g^{-1}R_{(i_2j_2})g^{-1}\cdots\right)
\end{eqnarray}
where $ R_{(ij)} $ is the matrix with the elements $ (R_{(ij)})_{kl} = R_{klij} $
and $g^{-1}$ has the elements $(g^{-1})_{km}=g^{km}$. This may be obtained from
%\begin{eqnarray}
$ F^{ab}_{ij} =  R_{klij}g^{km}e^a_m e^{bl}$.
%\end{eqnarray}

In 3 space dimensions, the Riemann curvature tensor can be completely expressed in terms of the
Einstein tensor:
\bea
R_{klij}=-g\,\epsilon_{klm}\epsilon_{ijn}G^{mn},    \label{18}
\eea
where $g={\rm det}g_{ij}$. 
This allows us to express the Yang-Mills action in 3d
\bea
S=\frac{1}{2g_{YM}^2}\int d^3x\,B_i^a B_i^a     \label{YM}
\eea
(where $B_i^a\equiv \frac{1}{2}\epsilon_{ijk}F_{jk}^a$) also in terms of the Einstein tensor, as follows.
In (\ref{riemann}), we use (\ref{18}) and also
$F^{ab}_{ij}=-\epsilon^{abc}F^c_{ij}$.
This gives \cite{hj}
\bea
B_i^a=({\rm det}\,e)e^{aj} G_{ij}                   \label{18'}
\eea
where ${\rm det}\,e={\rm det}e^a_i=\sqrt g$.
Then the Yang-Mills action in 3d is
\bea
S=\frac{1}{2g_{\rm YM}^2}\int d^3x\,g\,g^{jk}G_{ij}G_{ik}\,.
\eea
This is like $R^2$ gravity, but {\it without the diffeomorphism invariance}.

Eq.\ (\ref{db}) or (\ref{db2}) comprise a set of linear equations for the variables $ A^a_i $:
\begin{eqnarray}
\label{db1}
\epsilon_{ijk}(\d_j e^a_k + \epsilon^{abc} A^b_j  e^c_k) = 0.
\end{eqnarray}
Eq.\ (\ref{db1}) has diffeomorphism covariance provided both $A_i^a$ and $e_i^a$ transform as
covariant GL(3) vectors (the Christoffel symbol does not contribute due to antisymmetry of $\epsilon_{ijk}$).
Note that $\epsilon_{ijk}$ in (\ref{db1}) can be replaced by the Levi-Civita tensor 
$\varepsilon^{ijk}$ since the two differ by just a factor of $\sqrt g$.
% (see (\ref{B2}) below).

When the matrix $\{e_i^a\}$ is non-singular, (\ref{db1}) can be uniquely solved for $ A^a_i $.
Indeed, multiplying by $ e^a_l $ and summing over $ a $ we get
  \begin{eqnarray}
\epsilon_{ijk}e^a_l \d_j e^a_k = a_{il}-\delta_{il} a_{mm}           \label{eq12}
\end{eqnarray}
where $ a_{jm} = ({\rm det}\,e) A^b_je^{bm}$. Now set $i=l$ in (\ref{eq12}) to get $a_{mm}$ and put it back in
(\ref{eq12}). Therefore (\cite{hj}, \cite{pushan1})
\begin{eqnarray}
\label{A}
A^b_l = \frac{1}{{\rm det}\,e} \epsilon_{ijk} e^a_l \left(e^b_i\d_j e^a_k - \half \delta_{ab} e^c_i\d_j e^c_k\right)\,.
\end{eqnarray}

To sum up,
eq. (\ref{db}) or (\ref{db2}) or (\ref{db1}) defines $e^a_i$. In our context, the dreibein does not arise
from transformation between coordinate basis and orthonormal basis. The gauge transformation of $A^a_i$ ensures
that $e^a_i$ transforms homogeneously and so $g_{ij}$ as defined by (\ref{metric}) is gauge invariant.
Eq.\ (\ref{db1}) is diffeomorphism covariant. But the action is not
diffeomorphism invariant, since we are dealing with Yang-Mills theory in flat space and not in curved space.

%%%%%%%%%%%%%%%%%%%%%%%%%%%%%%%%%%%%%%%%%%%%%%%%%%%%%%%%%%%%%%%%%%%%%%%%%%%%%%%%%%%%%%%%%%%%%%%%%%%%%%%
\section{\label{sec:3}On existence and uniqueness of $e^a_i$ for given $A^a_i$} 
%%%%%%%%%%%%%%%%%%%%%%%%%%%%%%%%%%%%%%%%%%%%%%%%%%%%%%%%%%%%%%%%%%%%%%%%%%%%%%%%%%%%%%%%%%%%%%%%%%%%%%%
When $A^a_i(x) = 0$, equation (\ref{db1}) implies $e^a_i$ is curl-free for each $a = 1, 2, 3$
and we have a general solution
\begin{eqnarray}
\label{A1}
e^a_i(x) = \d_i\varphi^a(x)
\end{eqnarray}
where $\varphi^a(x)$ are arbitrary functions. Thus equation (\ref{db1}) has a large set of zero modes. 
In this situation, the curvature $F^{ab}_{ij}$  (or $R_{klij}$) vanishes and we have a flat space. 
From (\ref{metric}), we see that (\ref{A1}) 
 corresponds to  $g_{ij}(x) = \d_i \varphi^a\d_j\varphi^a$, i.e. a set of curvilinear coordinates
$\phi^a(x)$ of the flat space.
Now equation (\ref{db1}) 
is covariant under diffeomorphisms.
If we start with $A^a_i(x) = 0$ and equation (\ref{A1}), and apply the diffeomorphism $x\rightarrow x'(x)$
using the transformation property
of covariant vectors, $A^a_i$ stays zero but $e^a_i$ changes to ${e^a_i}'(x)=\partial_i\theta^a(x)$
(where $\theta^a(x')=\phi^a(x)$), i.e. we have a new dreibein for the same $A^a_i$.
Thus {\it we may view the set of zero modes of (\ref{db1}) given by (\ref{A1})
as accidental to the case of $ A^a_i=0$ and as a result of diffeomorphism covariance}.

In Ref.\ \cite{hj}, it is claimed that all pure gauges, and not just $ A^a_i=0$, give zero modes.
But a pure gauge potential $A_i(x)=O^T(x)\partial_i O(x)$ (where  
$A^{ab}_i =-\epsilon^{abc} A^c_i$ and $O$ is an SO(3) matrix) changes under both diffeomorphism and
gauge transformation. If $A_i^a$ were invariant but the dreibein changed under a transformation,
we would have had non-unique $e_i^a$ for a given pure gauge $A_i^a$. This does not appear to be the case.
So, while we do not exclude zero modes  for non-vanishing pure gauges,
such zero modes do not follow from any symmetry (either gauge covariance or diffeomorphism covariance) 
of equation
(\ref{db1}), unlike the zero modes for vanishing gauge potential.

Now {\it both vanishing gauge potential and pure gauge $A_i^a$ correspond to $B_i^a=0$.}
The point we make in this section is that for {\it a generic gauge potential with $ {\rm det} B_i^a \ne 0 $},
we do not have such large set of non-unique $e_i^a$.
{\it This issue was earlier addressed in Sec.\ II of Ref.\ \cite{pushan3}.
Here we present a simpler and neater analysis, arriving at the same conclusion.}

First we note that operating on (\ref{db2}) by $D_i$ gives a {\it consistency condition}, as follows.
We have $\epsilon_{ijk} D_i^{ca} D_j^{ab} e^b_k=0$. Using antisymmetry of $\epsilon_{ijk}$ and
$[D_i, D_j]^{cb}=F_{ij}^{cb}=-\epsilon^{cba}F_{ij}^a$, we arrive at 
a condition to be satisfied by any solution to (\ref{db1}):
\begin{eqnarray}
\label{A6}
\vec{B}_i(x)\times \vec{e}_i(x)= 0\,. 
\end{eqnarray}
 
{\it In the case $ {\rm det} B_i^a \ne 0 $}, without any loss of generality, we may expand 
$\vec{e}_i(x) $ in the  basis provided by $\vec{B}_i$ ($ i = 1, 2, 3 $): 
\begin{eqnarray}
\label{A9}
e^a_i(x) = B^a_j \alpha_{ji}.
\end{eqnarray}
Putting ({\ref{A9}) in the consistency condition (\ref{A6}) gives us 
\begin{eqnarray}
\label{A10}
({\rm det}\,B)\epsilon_{ijk}(B^{-1})^a_k \alpha_{ji} = 0.
\end{eqnarray}
Multiplying (\ref{A10}) by $B_l^a$, we
get $\epsilon_{ijl}\alpha_{ji}=0$.
Therefore $ \alpha_{ij}(x) $ is a symmetric matrix:
\begin{eqnarray}
\label{A11}
\alpha_{ij}(x) = \alpha_{ji}(x)
\end{eqnarray}
In terms of these variables $ \alpha_{ij}$,
the equations (\ref{db1}) become 
\begin{eqnarray}
\label{A12}
\epsilon_{ijk}\left(\d_j \alpha_{mk} + \beta_{mjl}\alpha_{lk}\right)  = 0
\end{eqnarray}
with $ i, j, k, l, m = 1, 2, 3  $.  Here
\begin{eqnarray}
\label{A13}
\beta_{mjl} = (B^{-1})_m^a (D_j B_l)^a\,.
\end{eqnarray}
(\ref{A12})  give the following six equations for evolution of $ \alpha_{ij}$ in $ x_3 $:
\begin{eqnarray}
\label{A14}
\d_3 \alpha_{11} - \d_1\alpha_{13} + \beta_{13l}\alpha_{l1} - \beta_{11l}\alpha_{l3} &=& 0\\
\label{A15}
\d_3 \alpha_{21} - \d_1\alpha_{23} + \beta_{23l}\alpha_{l1} - \beta_{21l}\alpha_{l3} &=& 0\\
\label{A16}
\d_3 \alpha_{31} - \d_1\alpha_{33} + \beta_{33l}\alpha_{l1} - \beta_{31l}\alpha_{l3} &=& 0\\
\label{A17}
\d_3 \alpha_{12} - \d_2\alpha_{13} + \beta_{13l}\alpha_{l2} - \beta_{12l}\alpha_{l3} &=& 0\\
\label{A18}
\d_3 \alpha_{22} - \d_2\alpha_{23} + \beta_{23l}\alpha_{l2} - \beta_{22l}\alpha_{l3} &=& 0\\
\label{A19}
\d_3 \alpha_{32} - \d_2\alpha_{33} + \beta_{33l}\alpha_{l2} - \beta_{32l}\alpha_{l3} &=& 0
\end{eqnarray}
Note that there is no evolution equation for the variable $ \alpha_{33} $. 
Also, since there are two evolution equations 
(\ref{A15}) and (\ref{A17}) for the same variable $ \alpha_{12} = \alpha_{21} $,
we get the consistency requirement
\begin{eqnarray}
\label{A20}
\d_3 \alpha_{12} = \d_2 \alpha_{13} + \beta_{12l}\alpha_{l3} - \beta_{13l}\alpha_{l2}
= \d_1 \alpha_{23} + \beta_{21l}\alpha_{l3} - \beta_{23l}\alpha_{l1}\,.
\end{eqnarray}
This has to be satisfied at all $ x_3 $. The variable $ \alpha_{33} $ enters in this consistency equation. If
the coefficient of $ \alpha_{33} $, that is, 
\begin{eqnarray}
\label{A21}
\beta_{123} - \beta_{213} \ne 0
\end{eqnarray}
 we can solve  for $ \alpha_{33} $ in favour of the other five variables $ \alpha_{11}, \alpha_{12}, \alpha_{13}, \alpha_{22}, \alpha_{23}$.
  Substituting this for $ \alpha_{33} $ in the  (five) independent equations 
(\ref{A14})-(\ref{A19}), we get evolution equations for these five independent
  variables. The solution is unique with initial choice on $ x_3 = x^0_3 $ surface.
This is ensured by the Cauchy-Kowalevski theorem. (See, for example, Ref.\ \cite{pushan3}, which
contains further reference, for a statement of this theorem.)
In the generic case, at least one of $(\beta_{123} - \beta_{213})$,
$(\beta_{231} - \beta_{321})$ and $(\beta_{312} - \beta_{132})$ will be non-zero.
If $\beta_{231} - \beta_{321}\neq 0$ (or $\beta_{312} - \beta_{132}\neq 0$),
 we can consider the evolution equation in
$x_1$ (or $x_2$) instead of $x_3$ and obtain unique solution with
initial choice on $ x_1 = x^0_1 $ (or $ x_2 = x^0_2 $) surface.

We have demonstrated that 
on fixing the boundary condition, a unique torsion-free dreibein $e_i^a$ exists for a
generic Yang-Mills potential $A_i^a$. So,
after taking into account the zero modes discussed in the
first two paragraphs of this section,
it should be possible to calculate the
Jacobian of the transformation.
Thus the transformation from $A_i^a$ to $e_i^a$ as given by (\ref{db}) is indeed possible.
However the Jacobian will not be needed in the calculations of this paper. Only the defining
equation of the dreibein will be used. 

%%%%%%%%%%%%%%%%%%%%%%%%%%%%%%%%%%%%%%%%%%%%%%%%%%%%%%%%%%%%%%%%%%%%%%%%%%%%%%%%%%%%%%%%%%%%%%%%%%%%%%%%
\section{\label{sec:4}Non-perturbative monopole configurations and conformal mode of metric}
%%%%%%%%%%%%%%%%%%%%%%%%%%%%%%%%%%%%%%%%%%%%%%%%%%%%%%%%%%%%%%%%%%%%%%%%%%%%%%%%%%%%%%%%%%%%%%%%%%%%%%%%
It is expected that topological degrees of freedom like magnetic monopoles are responsible for non-perturbative
properties such as  confinement \cite{mandel}. In the Georgi-Glashow model,confinement is due to
the 't Hooft-Polyakov monopole solution \cite{polyakov}. In contrast to this model,
there is no scalar (Higgs) field in the SO(3) Yang-Mills theory in three dimensions.
But the gauge field part of the 't Hooft-Polyakov monopole with appropriate boundary conditions  (see below)
constitutes a finite-action field configuration
and contributes to the functional integral. Moreover it has strong qualitative effects on the Wilson loop.
  
We now summarize the idea of {\it `centres' of topological field configurations}
as developed and used
in our earlier works \cite{pushan1,hms1,hms2}.
 (This framework is in the spirit of, but different from the Abelian projection procedure of 't Hooft \cite{hooft1}.)
The topological properties of the 't Hooft-Polyakov monopole  can be characterised using only
the gauge field and in the interior (in contrast to the usual characterisation by the triplet scalar field
at infinity). This is therefore useful in pure gauge theory.
This characterisation 
is done using the eigenvector fields of the
gauge invariant
quantity $S_{ij}\equiv B_i^{a}B_j^{a}$, where $B_i^a\equiv \frac{1}{2}\epsilon_{ijk}F_{jk}^a$  
is the non-Abelian magnetic field. 

For the 't Hooft-Polyakov monopole, $S_{ij}=\alpha(r)\delta_{ij}+\beta(r)x_i x_j$, where
$\alpha$ and $\beta$ are functions of the distance $r$ from the origin. At $r=0$, $S_{ij}\sim\delta_{ij}$
and so any direction is an eigenvector. Thus the eigenvector $x_i$ with unit winding number is singular at
the origin. In general,
the points at which the eigenvalues of $S_{ij}$ become 
triply degenerate,
or equivalently, the eigenvector fields of $S_{ij}$ become singular, 
are called the `centres' of the monopoles and other topological objects in Yang-Mills theory.

Let us then consider the gauge field part of the 't Hooft-Polyakov monopole configuration
\bea
A_i^a(x) = \epsilon _{aij} x_j \frac{1-K(r)}{r^2}      \label{tHP}
\eea
Here $K(r)=1+O(r^2)$ for $r\rightarrow 0$ and $K(r) \rightarrow 0$
for $r\rightarrow \infty$. (These conditions ensure the finiteness of the action.) Using
\bea
e^a_i(x)=\delta^a_i e^{-g(r)}       \label{g(r)}
\eea
in  (\ref{A}), we find that the resulting gauge potential corresponds to (\ref{tHP}) with
\bea
\frac{dg}{dr}=\frac{1-K(r)}{r}\,.    \label{dgdr}
\eea

The configurations (\ref{tHP}) and (\ref{g(r)}) possess spherical symmetry. 
We next consider a generalization of (\ref{g(r)}) with $\phi(x)$,
an arbitrary scalar function of $x_1$, $x_2$ and $x_3$, in the place of $g(r)$:
\bea
e^a_i(x)=\delta^a_i e^{-\phi(x)}\,.                       \label{ansatz}
\eea
This corresponds to 
\bea
g_{ij}=\delta_{ij}e^{-2\phi}\,,                          \label{gphi}
\eea
a conformally flat metric, and we call $\phi$ the {\it conformal mode}.
Using (\ref{ansatz}) in (\ref{A}) now leads to 
\bea
A_i^a(x) = \epsilon _{aij}\partial_j\phi\,.
\eea
This gives 
\bea
B_i^{a}=\partial_i\partial_a\phi-\delta_{ia}\partial^2\phi+\partial_i\phi\partial_a\phi\,.
\label{Bia}
\eea
Now {\it consider a point at which the following hold}:
\begin{enumerate}
\item
$\partial_i\phi=0$.
\item
$\partial_i\partial_a\phi=0$ for $i\neq a$.
\item
$\partial^2\phi/\partial x_1^2=\partial^2\phi/\partial x_2^2=\partial^2\phi/\partial x_3^2$.
\end{enumerate}
Note that the first condition says that we have a critical point and the next two conditions say that
the Hessian matrix is diagonal and its three diagonal elements (eigenvalues) are equal. These are
sufficient conditions for $\phi$ to be an extremum at the point. ($\phi$ attains a maximum
or a minimum depending on whether the eigenvalues are all positive or all negative.)

Now since the conditions 2 and 3 are equivalent to 
$\partial_i\partial_a\phi=\frac{1}{3}\delta_{ia}\partial^2\phi$, it follows from (\ref{Bia}) that
at the point under consideration,
$B_i^{a}\sim \delta_{ia}$ and so $B_i^{a}B_j^{a}\sim \delta_{ij}$. 
So the matrix $S_{ij}$ is triply degenerate at the
point, which, as explained at the beginning of this section, 
is the criterion for locating the `centre' of a topological configuration in
a gauge-invariant description. Thus {\it at the topological centre, the field $\phi$ is
an extremum, satisfying the three conditions given above.}

For the special, spherical symmetric case given by (\ref{tHP}) and (\ref{g(r)}), 
using $K(r)=1+O(r^2)$ for $r\rightarrow 0$,
it can be checked that the function $g(r)$ indeed satisfies the above conditions on $\phi(x)$ 
at the monopole centre $r=0$.
On the other hand, in the more general
form given by (\ref{ansatz}),  
the conditions on $\phi(x_1, x_2, x_3)$ stated above show that 
$\phi$ needs to be symmetric only upto the second order
terms in the Taylor expansion about the centre. Thus, taking the topological centre to be at $r=0$,
the forms $A+Br^2+Cx_1^3+\cdots$ and $A+Br^2+Dx_1^2x_2+\cdots$ (where $A$, $B$, $C$, $D$ are constants)
are two examples of possible Taylor expansion for the function $\phi$ satisfying the conditions.

It is interesting to apply our formalism to the Wu-Yang monopole \cite{wy}. This is a solution
to the Yang-Mills field equations. However it is a point-like magnetic monopole. The gauge potential
for it is obtained on putting
$K(r)=0$ in (\ref{tHP}). Then (\ref{dgdr}) gives $g(r)=\ln r$ (upto an additive constant, which sets the scale).
So the conformal mode goes to negative infinity as $r$ goes to zero. Such a singularity, instead of a smooth minimum,
in the conformal mode is just what is expected since the Wu-Yang monopole is point-like.
The metric in this case is $g_{ij}=\delta_{ij}/r^2$ (upto a multiplicative constant).
  
%Using (\ref{Bia}), the Yang-Mills action of (\ref{YM}) becomes
%\bea
%S=\frac{1}{2g_{YM}^2}\int d^3x\,
%\Big[2(\partial^2\phi)^2- 3 \Big(\sum_i (\partial_i\phi)^2 \Big) \partial^2\phi + 
%\Big(\sum_i (\partial_i\phi)^2 \Big)^2\Big]
%\eea
%on integration by parts. 
%This massless scalar theory with quartic derivatives can
%serve as a toy model for understanding confinement by simulations.

%To evaluate the Jacobian $|\delta A^a_i(x)/\delta e^b_j(y)|$ for the case of the ansatz (\ref{ansatz}),
%we put (\ref{gphi}) in the affine connections in (\ref{Jfinal}). This gives
%\bea
%&&\int \D \bar{\eta}^{ai} \D \eta^b_j \exp\Bigg[-\int d^3x\, e^{-3\phi} \varepsilon^{ijk}\nonumber\\
%                &&\times\Big(\bar\eta_{il}\partial_j{\eta^l}_k
%                -e^{-4\phi}(\bar\eta_{il}{\eta^l}_k \partial_j\phi+\bar\eta_{ij}{\eta^l}_k \partial_l\phi
%                  -\bar\eta_{il}\eta_{jk} \partial^l\phi)          
%                    -\frac{1}{2}\bar{\eta^l}_l\partial_j\eta_{ik}\Big)\Bigg]\,.
%\eea

%%%%%%%%%%%%%%%%%%%%%%%%%%%%%%%%%%%%%%%%%%%%%%%%%%%%%%%%%%%%%%%%%%%%%%%%%%%%%%%%%%%%%%%%%%%%%%%%%%%%%%%
\section{\label{sec:5}Ansatz for gauge field}
%%%%%%%%%%%%%%%%%%%%%%%%%%%%%%%%%%%%%%%%%%%%%%%%%%%%%%%%%%%%%%%%%%%%%%%%%%%%%%%%%%%%%%%%%%%%%%%%%%%%%%%
We now consider the polar decomposition of the most general 3$\times$3 matrix $e_i^a$ into
an orthogonal matrix $\cal R$ and a symmetric matrix $\cal E$:
\bea
e_i^a(x)={\cal R}^a_j(x){\cal E}_{ji}(x)        ~~\forall x.                              \label{RE}
\eea
Under a gauge transformation,
\bea
{\cal R}^a_j(x)\rightarrow O^{ab}(x){\cal R}^b_j(x)\,,
\eea
while the symmetric matrix ${\cal E}_{ji}$ is gauge-invariant and is the symmetric square-root of the metric
$g_{ij}$. The decomposition (\ref{RE}) corresponds to
\bea
A_l^d(x)={\cal R}_m^d(x)a_{ml}(x)+\omega_l^d(x)\,,                                   \label{AA1}
\eea
where
\bea
a_{ml}(x)=\frac{1}{{\rm det}\,{\cal E}}
            \epsilon_{ijk}\left({\cal E}_{im}{\cal E}_{lp}-\frac{1}{2}{\cal E}_{ip}{\cal E}_{lm}\right)
                          \partial_j {\cal E}_{kp}
\label{AA2}
\eea
and $\omega_l^d$ is formally a pure gauge:
\bea
\omega_l^d(x)=-\frac{1}{2}\epsilon^{bcd}{\cal R}_m^b(x)\partial_l {\cal R}_m^c(x)\,.          \label{AA3}
\eea
(To obtain (\ref{AA1}), one puts (\ref{RE}) in (\ref{A}). The terms containing derivative of
$\cal E$ immediately give the first term on the RHS of (\ref{AA1}). In the terms containing derivative of
$\cal R$, we put $\partial_j{\cal R}_p^a=-\epsilon^{abc} \omega_j^b {\cal R}_p^c$,
which is the same as (\ref{AA3}). Then using 
$\epsilon^{abc}{\cal R}^a_n {\cal R}^c_p=\epsilon_{nqp}{\cal R}^b_q$ (since ${\rm det}\,R=1$)
and $\epsilon_{ijk}{\cal E}_{in}{\cal E}_{kp}=\epsilon_{nrp}({\rm det}\,{\cal E})({\cal E}^{-1})_{jr}$,
we get the $\omega_l^d$ term in (\ref{AA1}).)

Now the symmetric matrix ${\cal E}$ can be further decomposed into a spin-two traceless part and
a scalar trace part:
\bea
{\cal E}_{ij}=\tilde{\cal E}_{ij}+\delta_{ij}e^{-\phi}\,.                          \label{AA4}
\eea 
(Here $\tilde{\cal E}_{ij}={\cal E}_{ij}-\frac{1}{3}\delta_{ij}\Sigma_k{\cal E}_{kk}$
and we define $\frac{1}{3}\Sigma_k{\cal E}_{kk}$ to be  $e^{-\phi}$.)
{\it Since $\delta_{ij}e^{-\phi}$ corresponds to the metric (\ref{gphi}), we propose  that the topological 
configurations are contained in this part, with $\phi$ satisfying the three conditions given after
(\ref{Bia}) at isolated points.}
At such points, $\omega_l^d$
is not strictly a pure gauge. For example, the non-Abelian magnetic field
corresponding to $\omega_l^d$
has a Dirac string contribution when $\cal R$ is the singular gauge in which
the configuration (\ref{tHP}) becomes a Dirac monopole \cite{hms1}.

Thus equations (\ref{AA1}), (\ref{AA2}),(\ref{AA3}) and (\ref{AA4}) comprise our ansatz
for the gauge potential, which contains the topological aspects in a natural way.
The three gauge degrees of freedom reside in $\cal R$ and the six gauge-invariant degrees of freedom 
reside in $\cal E$. The conformal mode contained in  $\cal E$ captures the topological 
configurations.

%%%%%%%%%%%%%%%%%%%%%%%%%%%%%%%%%%%%%%%%%%%%%%%%%%%%%%%%%%%%%%%%%%%%%%%%%%%%%%%%%%%%%%%%%%%%%%%%%%%%%%%%%%%%
\section{\label{sec:6}Discussion}
%%%%%%%%%%%%%%%%%%%%%%%%%%%%%%%%%%%%%%%%%%%%%%%%%%%%%%%%%%%%%%%%%%%%%%%%%%%%%%%%%%%%%%%%%%%%%%%%%%%%%%%%%%%%
In this work, we have used the dreibein and the metric as basic variables for the SO(3) Yang-Mills
theory in three Euclidean dimensions.
The usefulness of this lies in the fact that the metric is gauge invariant.

By showing that a {\it generic} $A_i^a$ corresponds to a unique torsion-free $e_i^a$ on fixing
the boundary condition, we have provided justification
for the change of variables from $A_i^a$ to $e_i^a$.  

We have applied this formalism to non-perturbative aspects of the theory. The conformal mode of the metric
is found to contain the topological aspects of the gauge potential. 
We have demonstrated that  topological centres
 are located at certain points at which this conformal mode,
a scalar field, is an extremum. This criterion is thus an alternative to that developed by us in
Refs.\ \cite{pushan1} and \cite{hms1} (also applied by us in \cite{hms2}), namely, that
the points of triple degeneracy
of $B_i^a B_j^a$ are the centres of topological configurations.
Both of these criteria are gauge invariant.
The Yang-Mills theory expressed in terms of this conformal mode, would be a theory
of a scalar field, and lattice simulation of this {\it scalar} theory (as opposed to lattice {\it gauge} theory)
with monopoles located at extrema of the scalar field, would serve as a model for confinement.
 
We have also proposed an ansatz for the gauge potential, which incorporates the topological aspects
contained in the conformal mode after separating the gauge variant and the gauge invariant degrees of freedom. 

When our formalism is applied to the Hamiltonian formulation of 3+1-
dimensional Yang-Mills theory, the canonical variables are the metric and its conjugate variable.
The physical wavefunctionals are simply functionals of the metric.
However, due to the change of variables, the inner product of the wavefunctionals
will involve the Jacobian of the transformation.
%\bea
%d\mu(A)=|\delta A/\delta e| d\mu(e), \\
%\langle 1|2\rangle=\int d\mu(e) |\delta A/\delta e| \Psi_1^*\Psi_2.
%\eea
For 2+1-dimensional SU($N$) gauge theory,  Nair et al (\cite{nair1}-\cite{nair4})
proposed an $N\times N$ complex matrix as a prepotential, and, as  a consequence of the change of variables,
obtained an inner product for physical
states which involves the Wess-Zumino-Witten action.
They arrived at
a trial wave functional which exhibits confinement and gives string tension in striking
agreement with lattice gauge theory simulations. 
Viewing our proposal of using the dreibein as prepotential as a generalization to 3+1-dimensions, 
it will therefore be interesting to calculate the Jacobian of the transformation in our case.

\section*{Acknowledgement}

I.M. thanks UGC (DRS) for support.

\section*{Competing Interests}

The authors declare that there is no conflict of interest regarding the publication of this paper.

\end{document}